\documentstyle[aps,prb,epsfig,floats]{revtex}

\begin{document}
%\draft
%\title{Effects of crack tip geometry on dislocation emission and
%  cleavage: A possible path to enhanced ductility}
%\author{J. Schi{\o}tz\cite{JSaddr}, L. M. Canel, and A. E. Carlsson}
%\address{Department of Physics, Washington University, St.\ Louis, MO
%  63130-4899, USA}
%\date{\today}
%\maketitle

\twocolumn[
\begin{center}
  \large\bf Effects of crack tip geometry on dislocation emission and
  cleavage: A possible path to enhanced ductility
\end{center}
\begin{center}
  J. Schi{\o}tz\cite{JSaddr}, L. M. Canel, and A. E. Carlsson\\
  {\em Department of Physics, Washington University, St.\ Louis, MO
    63130-4899, USA}\\
  (June 27, 1996)
\end{center}

\begin{quote}
\begin{quote}\setlength{\parindent}{1em}
%\begin{abstract}
  \hspace*{1em}%
  We present a systematic study of the effect of crack blunting on
  subsequent crack propagation and dislocation emission.  We show that
  the stress intensity factor required to propagate the crack is
  increased as the crack is blunted by up to thirteen atomic layers,
  but only by a relatively modest amount for a crack with a sharp
  60$^\circ$ corner.  The effect of the blunting is far less than 
  would be expected from a smoothly blunted crack; the sharp corners
  preserve the stress concentration, reducing the effect of the
  blunting.  However, for some material parameters blunting changes
  the preferred deformation mode from brittle cleavage to dislocation
  emission.  In such materials, the absorption of preexisting
  dislocations by the crack tip can cause the crack tip to be locally
  arrested, causing a significant increase in the microscopic
  toughness of the crack tip.  Continuum plasticity models have shown
  that even a moderate increase in the microscopic toughness can lead
  to an increase in the macroscopic fracture toughness of the material
  by several orders of magnitude.  We thus propose an
  atomic-scale mechanism at the crack tip, that ultimately may lead to
  a high fracture toughness in some materials where a sharp crack
  would seem to be able to propagate in a brittle manner.
  
  When the crack is loaded in mode II, the load required to emit a
  dislocation is affected to a much higher degree by the blunting, in
  agreement with the estimates from continuum elasticity.  In mode II
  the emission process is aided by a reduction of the free surface
  area during the emission process.  This leads to emission at crack
  loadings which are lower than predicted from the continuum analysis
  of Rice.
%\end{abstract}
\end{quote}
\end{quote}

%\pacs{62.20.Mk, 61.72.Lk, 61.62.Yk}
\begin{center}
  (Submitted to {\em Phys.\ Rev. B})
\end{center}

]

%\narrowtext

\section{Introduction}
\label{sec:intro}

The ability to calculate, or even to understand quantitatively, the
fracture toughness of materials has long been one of the major goals
of materials science.  Unfortunately, that goal has yet to be
realized.  The phenomena that cause some materials to be brittle and
some to be ductile involve mechanisms at vastly different length
scales, from the atomic-scale processes at crack tips and dislocation
cores, to the formation of cell structures and texture at much larger
length scales.

Recently, the atomic-scale behavior of the crack tip has been modeled
with some success.  Generally, the behavior of cracks falls onto two
categories.  In ``intrinsically brittle'' materials, an atomically
sharp crack will propagate, possibly leading to cleavage of the
specimen.  In ``intrinsically ductile'' materials, the crack will emit
one or more dislocations instead of propagating, and brittle fracture
becomes impossible.  Significant progress has been made in
understanding the intrinsic behavior of the crack tip in terms of well
defined atomic-scale
energies\cite{RiTh74,LiTh86,Sc91,Ri92,ZhCaTh94,ShAn94,Gu95}.

The intrinsic behavior of the crack tip does not by itself determine
the behavior of the material.  Most materials contain a density of
dislocations and dislocation sources that is high enough to affect
fracture strongly.  It is perfectly possible that an ``intrinsically
brittle'' material may behave in a ductile manner, if for
example a loaded crack causes dislocations to be emitted from nearby
sources, which subsequently shield the crack.  Modeling of
these processes is impractical at the atomic scale, but requires the
use of continuum or quasi-continuum
methods\cite{ZhTh91,KuCaCoDePoBr92,BaCa95,ZaSrLe96}.

Even if the dislocation activity in the material surrounding a crack
is in itself insufficient to change the behavior, the dislocations may
directly interact with the crack tip, and thus change its behavior.
If the crack tip intersects a dislocation a localized step or jog in
the crack tip is formed.  The jog may act as a local nucleation site
for dislocations, favoring ductile behavior\cite{ZhTh91b,ArXuOr96}.
But even without introducing inhomogeneities the detailed atomic
configuration may significantly change the behavior of the crack: In a
recent paper\cite{ScCaCaTh96} we demonstrated that the emission of
dislocations becomes favored if the crack tip is blunt at the atomic
level.  In this paper this effect is investigated further.  This
supplements earlier work showing that the effect of blunting on crack
propagation is smaller than what one would expect from relatively
simple continuum elasticity considerations\cite{Gu95,PaMaShSiDi85}. We
present arguments for this reduced effect based on a more detailed
continuum model that includes the sharp corners of the crack.  In
addition, simple models of the inter-atomic interactions permit us to
analyze the results in a more detailed way, and to understand the
effects of blunting as a combination of linear elastic and nonlinear
phenomena.

The effect of crack blunting on crack behavior may be of significant
practical interest in some materials.  Blunting may occur due to
emission from the tip or by collision of the tip with pre-existing
dislocations.  If this blunting is enough to arrest the crack the
macroscopic fracture toughness of the material may change significantly.  The
absorption of pre-existing dislocations may be enhanced by attractive elastic
interaction between the crack tip and the dislocations.  Recent
simulations\cite{Me96,Me96b} have shown that the stress field of a
crack moving through a material will cause dislocations located within
the immediate vicinity of the crack path to be attracted to the
crack.  Since dislocations lying within several hundred lattice
constants of the crack path may collide with the tip, the likelihood
of blunting due to intersection of the crack tip with pre-existing
dislocations is high.

We have also studied the behavior of the blunt crack under mode II
loading.  Here we find a much larger effect of the blunting on the
load required to emit a dislocation, in agreement with arguments from
continuum elasticity.  We also find that the surface energy plays a
prominent role in the emission process.  Emission under mode II
conditions is observed to be accompanied by a reduction in the total
surface area.  When the surface energy is high, emission occurs at
loadings which are much lower than the corresponding values predicted
by Rice\cite{Ri92}.

In the next section we present the method used for our simulations.
In section \ref{sec:modeI} we present the results for a blunt crack
loaded in mode I.  We find that blunting the crack tip results in a
modest increase in the load required to propagate it, and that often
the crack response changes from brittle cleavage to dislocation
emission.  We discuss how this may influence the fracture toughness of
a metal.  Section \ref{sec:modeII} is a brief overview of the results
in mode II loading, where we find that the surface energy plays a
prominent role in the emission process.  Finally, we present
analytical calculations on stress fields around a blunt crack in the
Appendix.

\section{Simulation methods.}
\label{sec:simmethods}

The slow decay of the elastic fields around a crack ($\sigma \sim
r^{-1/2}$) makes atomic-scale simulations of cracks very challenging,
since large system sizes are needed to minimize the effect of the
boundary conditions.  Many methods have been used to get around these
problems, such as embedding the atomistic simulation in a continuum
model \cite{KoGuFi91,TaOrPh96} or using extremely large systems on
state-of-the-art parallel computers\cite{ZhLoThHo96}.

%%%%%%
%
%  FIGURE 1
%
%%%%%%
\begin{figure}[t]
  \begin{center}
    \epsfig{file=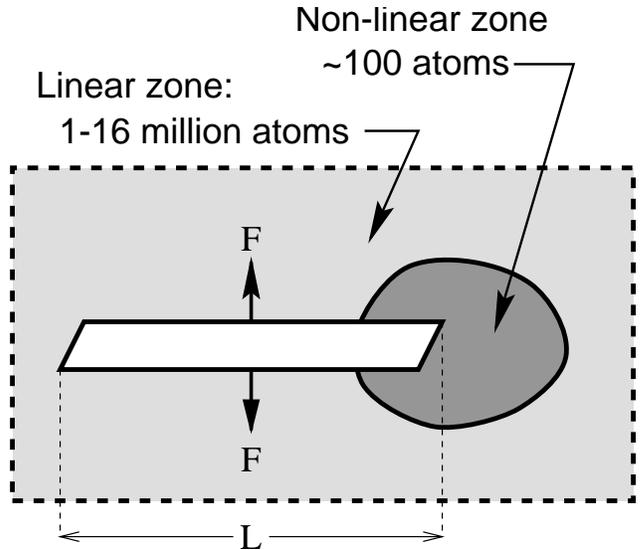,clip=}
    \vspace{2mm}
    \caption{Division of the system into a large linear zone and a
      small nonlinear zone.  Atoms in the linear zone are treated
      indirectly through the use of a Green's function.  The
      non-linear zone is placed around one end of the crack.  The
      crack is loaded in the middle by a force dipole, and its length
      $L$ is kept constant at 65 lattice spacings as the blunting is
      varied.}
    \label{fig:globalview}
  \end{center}
\end{figure}
Our solution is to model the elastic response of the surrounding
medium by a Green's function.  The simulation cell is divided into two
zones (see figure \ref{fig:globalview}). The atoms located near the
crack tip interact with each other through a non-linear force law (the
{\em non-linear zone\/}), whereas the atoms far from the crack tip
interact only through linear forces.  The displacements of the atoms
in this {\em linear zone\/} can then be fully described by a lattice
Green's function $G_{ij}({\bf r}, {\bf r}')$, which describes the
displacement of the atom at ${\bf r}$ to a force acting on the atom at
${\bf r}'$, taking the presence of the crack into account.  The
interactions of atoms in the non-linear zone with atoms in the linear
zone are also described by the Green's function, and are therefore
linear.  The Green's function can be calculated in a computationally
efficient way\cite{ThZhCaTe92}.  The total energy of the system can
then be described as the sum of the energy in the elastic far field
(calculated from the Green's function) and in the non-linear
interactions\cite{CaCaTh95}.  This typically reduces the number of
degrees of freedom in the problem from approximately $10^7$ to
$10^2$--$10^3$.

We use this method to study the deformation modes of blunt cracks with
up to thirteen atomic layers of blunting, for a range of force laws.
Figure \ref{fig:modes}a shows a typical initial configuration.  We
have a region in front of the crack where the atoms are allowed to
move freely.  Two ``spurs'', along which dislocations generated at
the crack tip can move away from the crack, are included in the
non-linear zone.  Since bonds cannot be broken and reformed in the
linear zones, dislocations will be unable to leave the non-linear
zone.  The two-dimensional hexagonal lattice studied contains three
slip planes at $60^\circ$ angles.  The crack is oriented parallel to
one of the slip planes.

The inter-atomic interactions in the non-linear zone are described by
the UBER pair potential\cite{RoSmFe83}:
\begin{equation}
  \label{eq:uber}
  F(r) = -k (r - r_0) \exp \left({r - r_0 \over l} \right)
\end{equation}
where $r$ is the inter-atomic separation, $r_0$ is the separation in
equilibrium, and $l$ is a range parameter.  The interactions are cut
off so that only nearest-neighbor interactions are included, and
forces are shifted slightly to avoid steps in the force at the cutoff.
Further, a small scaling $C$ of the force law is used to preserve the
elastic constants, thus enabling us to use the same Green's function
for all the force laws.  The force law thus becomes
\begin{equation}
  \label{eq:scaleduber}
  F(r) = C \left[ -k (r - r_0) \exp \left({r - r_0 \over l} \right) -
    F_0 \left({r - r_0 \over r_{\text{cut}} - r_0}\right) \right]
\end{equation}
where $F_0 = -k (r_{\text{cut}} - r_0) \exp
\left((r_{\text{cut}} - r_0) / l \right)$ assures that the force is
zero at the cutoff distance $r_{\text{cut}}$ (in this work $1.7
r_0$).  Since we do not study specific materials, no attempt
is made to use realistic many-body potentials.  In spite of the
relatively simple potentials used, this methodology is known to give
results that agree well with the observed behavior of real materials,
and with continuum models\cite{ZhCaTh94,ZhCaTh93}.  We also see good
agreement between our calculations and calculations using the Embedded
Atom Method\cite{Gu95,Lu95}, as discussed in section
\ref{sec:discussion}.

For all the calculations presented in this paper, the total system
size has been $1024 \times 1024$ atoms.  In order to establish that
the results are not influenced by the boundary conditions, we have
repeated a few of the simulations with 16 times as many atoms ($4096
\times 4096$).  The only difference is a {\em consistent\/} 1--1.5\%
reduction of the loads required to cleave the crack or to emit a
dislocation.  All trends remain the same.  The size-dependence is
caused by the stress fields of the cracks in neighboring supercells
causing the true stress intensity factor to be slightly different from
the calculated values.  Since this interaction remains the same it
does not influence our results, where we study the changes in the
critical loads.  We have also verified that increasing the size of the
non-linear zone does not affect the results.

\section{The blunt crack under mode I loading.}
\label{sec:modeI}

\subsection{Simulations.}
\label{sec:assymsim}

%%%%%%%%%
%
% FIGURE 2
%
%%%%%%%%%
\begin{figure}[t]
  \begin{center}
    \epsfig{file=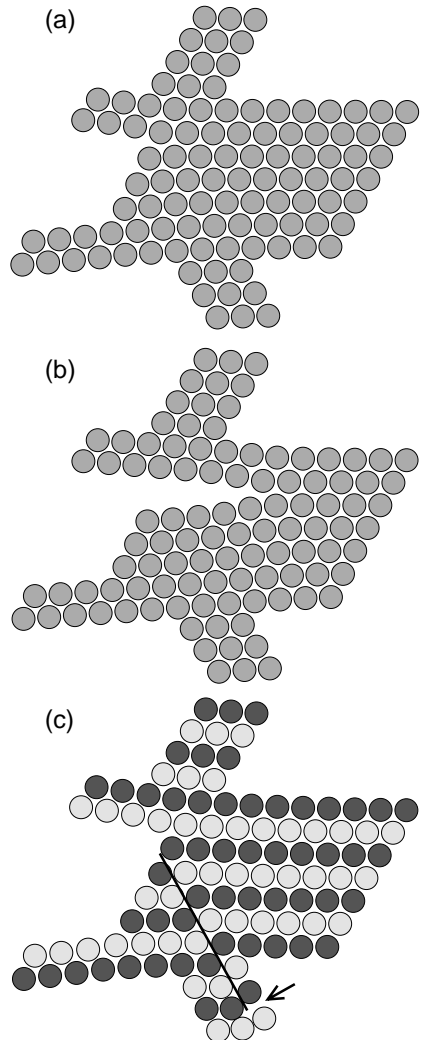,clip=}
    \caption{The two instability modes of the crack.  (a) shows the
      configuration before plastic deformation.  The crack, three
      atomic layers thick, extends out of the left of the picture.
      Only atoms in the non-linear zone are shown, see text.  (b)
      shows the crack propagating by cleavage.  (c) shows the crack
      emitting a dislocation.  The dislocation has traveled along the
      black line.  Since it cannot leave the non-linear zone, it has
      been pinned at the bottom (indicated by the arrow).  To make the
      path of the dislocation visible, the atoms were given two
      different shades.  Prior to emission the atoms were lined up in
      rows of the same shading.  Where the dislocation has traveled
      lines of atoms of different shade meet.}
    \label{fig:modes}
  \end{center}
\end{figure}
When we load the cracks until an instability occurs, one of two
different responses are observed: cleavage and dislocation emission
(figure \ref{fig:modes}).  The dislocation emission always occurs in
the downwards direction, as shown in the figure, except for the case
of a single layer of blunting, where the crack geometry is symmetric.
Furthermore, the emission preserves the shape of the blunted crack.
This means that further dislocations can be expected to be emitted in
the same direction, provided that the dislocation can move so far away
that its effect on the local stress field at the crack tip is small.
The crack shape studied is thus preserved during the blunting process.

%%%%%%%%%
%
% FIGURE 3
%
%%%%%%%%%
\begin{figure}[t]
  \begin{center}
    \epsfig{file=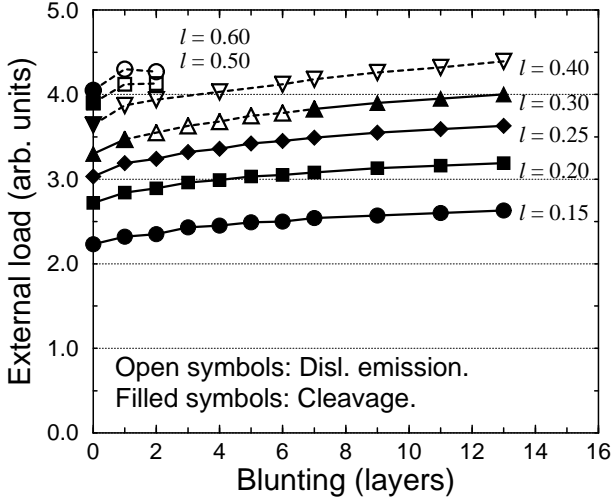,clip=}
    \vspace{2mm}
    \caption{The external load required for cleavage or dislocation emission
      as a function of the blunting, for seven different force laws.
      Open symbols indicate dislocation emission, closed symbols
      indicate cracking.  It is seen that for a wide range of force
      law parameters the sharp crack propagate by cleavage, whereas
      the blunt crack emits dislocations.  For $l \ge 0.5$ the
      crystal structure ahead of the crack changed from triangular
      to square under influence of the stress, making further
      simulations meaningless.}
    \label{fig:numresults}
  \end{center}
\end{figure}
Figure \ref{fig:numresults} shows the force required for cleavage or
dislocation emission as a function of the blunting, for seven
different decay lengths $l$ of the force law.  The results for the
sharp crack (zero blunting) are reported on the y-axis.  We find that
increasing the decay length $l$ increases the load required to
propagate the crack (see also Zhou {\em et al.\/}\cite{ZhCaTh94}).
This is because increasing $l$ raises the surface energy and therefore
the loading to cleave.  The critical values observed are 4--7\% above
the values expected from the Griffith criterion, this discrepancy is
attributed to the discreteness of the lattice (lattice
trapping\cite{ThHsRa71}).  Two main effects are seen with increasing
blunting.  For all force laws, blunting increases the critical
loadings to cleave or to emit, but only by a modest amount.  For 13
layers of blunting the increase is only approximately 20\%.  This
should be compared to estimates derived from linear elasticity of
elliptic cracks giving an effect of around 40\% for a single layer of
blunting, and 250\% for 13 layers\cite{Gu95}.  Secondly, for a range
of force law parameters $l$ a normally brittle crack is observed to
emit dislocations after being blunted by one or two layers.  Here,
although the force required to cause the blunt crack to emit is still
larger than the load that cleaves the sharp crack, the force required
to propagate the crack has increased even more, shifting the balance
between emission and cleavage.

%%%%%%%
%
% FIGURE 4
%
%%%%%%%
\begin{figure}[t]
  \begin{center}
    \epsfig{file=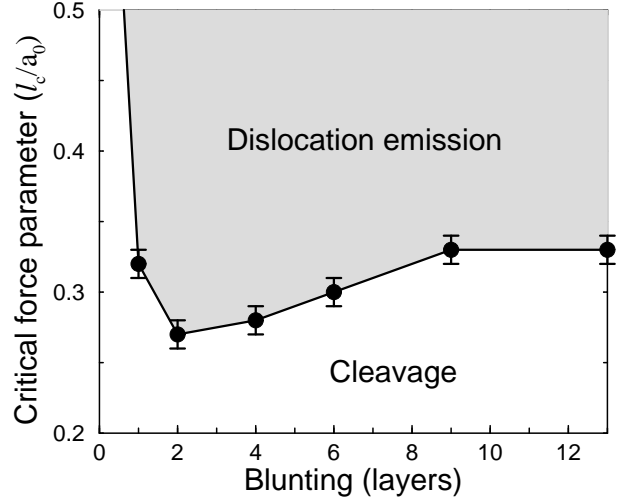,clip=}
    \vspace{2mm}
    \caption{The force law range parameter $l_c$ where cleavage and
      dislocation emission are in equilibrium, as a function of the
      blunting.  The first layer of blunting has a large effect,
      making a range of force laws ($l_c \ge 0.27$)
      emitting, further blunting makes some of these cleave again.}
    \label{fig:defmode}
  \end{center}
\end{figure}
For each value of the blunting, there is a transition from cleavage to
dislocation emission with increasing $l$, caused by an increase in
the surface energy.  Figure \ref{fig:defmode} shows the critical value
of the force law parameter $l_c$ for which the  crack response
changes from cleavage to emission.  The first layer of blunting
results in a significant reduction of $l_c$, causing a wide range
of force laws to favor emission.  Further increase in the blunting
causes $l_c$ to increase marginally.  Thus the first
layer or two of blunting tend to enhance ductility, and the subsequent
effects are much smaller.  We have not been able to explain this
effect on the basis of analytic theory.

\subsection{Discussion of the mode I results}
\label{sec:discussion}

%%%%%%%
%
% FIGURE 5
%
%%%%%%%
\begin{figure}[t]
  \begin{center}
    \epsfig{file=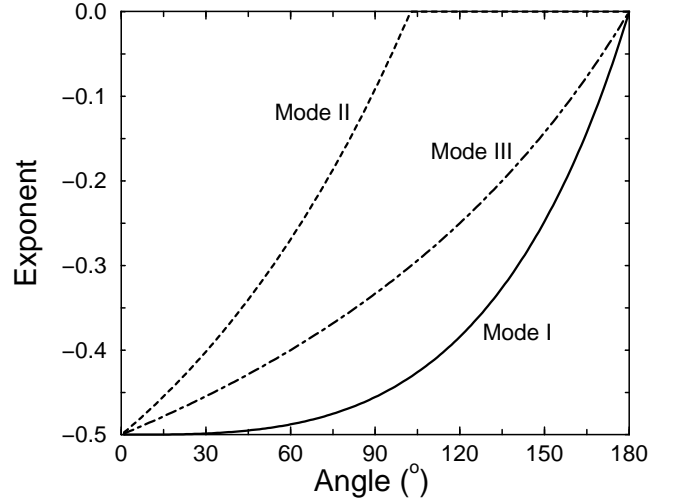,clip=}
    \vspace{2mm}
    \caption{The exponent of the stress singularity as
      a function of the opening angle for three different loadings.
      See the Appendix.}
    \label{fig:exponent}
  \end{center}
\end{figure}
%%%%%%%
%
% FIGURE 6
%
%%%%%%%
\begin{figure}[t]
  \begin{center}
    \epsfig{file=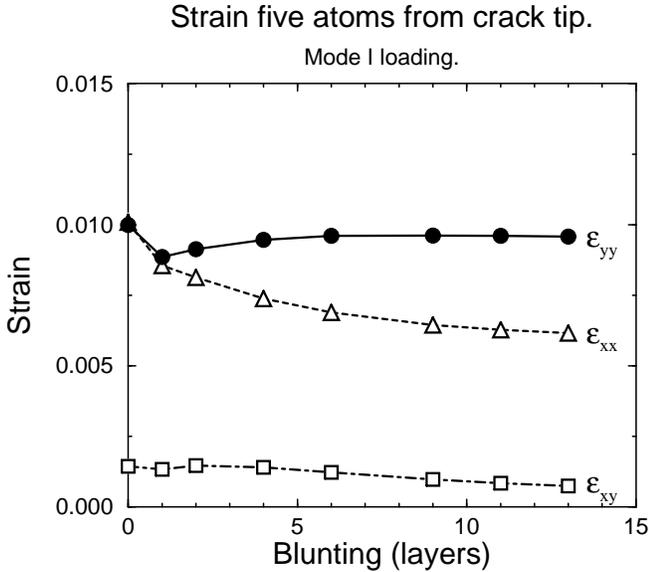,clip=}
    \vspace{2mm}
    \caption{The components of the elastic strain five atoms ahead of
      the crack as a function of blunting.  The loading of the crack
      is kept constant.}
    \label{fig:elaststress}
  \end{center}
\end{figure}
In order to understand these results, we have analyzed the nature of
the elastic fields around a $60^\circ$ wedge crack (see the Appendix).
Simplistically, one would expect that the increased opening angle
should reduce the stress singularity at the crack tip, making cleavage
more difficult.  However, we find that the strength of the stress
singularity at the end of a blunt crack with a $60^\circ$ angle is not
changed significantly compared to a sharp crack.  In particular, the
exponent of the singularity has only changed from \mbox{-0.5} to
\mbox{-0.488}, see figure \ref{fig:exponent}.  Therefore, purely from
considerations of the crack tip field, we would not expect any
significant change in the load required to propagate the crack, and
believe that this explains the smallness of the blunting effect.  This
expectation is confirmed by figure \ref{fig:elaststress}, which show the
elastic strain field five atoms ahead of the crack for a completely
linear (but still discrete) system.  The effect of the blunting on the
tensile strain ($\varepsilon_{yy}$), which should be the most important for
cleavage, is minimal, although some reduction is seen when the first
layer of blunting is introduced.  When the crack tip is blunted only
by one layer, the crack is symmetric and it does not resemble the shapes
studied in the Appendix; the behavior near a $60^\circ$
corner is thus not relevant, nor is continuum theory expected to be a
good description when only a single atomic layer has been removed.  As
the blunting is increased, the actual crack shape approaches that of
the ideal theoretical shape studied in the Appendix, and the stress
field then approaches the value corresponding to the wedge crack.
Since that value is virtually indistiguishable from the stress in
front of a sharp crack, the stress is seen to increase again when the
blunting increases.

The decrease in strain when the first layer of blunting is introduced
is consistent with the observed increase in the force required to
propagate the crack.  However, the elastic behavior completely fails
to explain the additional increase in the force to cleave as the crack
is further blunted, since it would indicate that this force should
decrease slightly again\cite{Endnote1}.
%\footnote{The change in the exponent from
%  $-1/2$ to $-0.488$ would indicate that the strain five atoms ahead
%  of the crack should have decreased by approximately 1\% compared to
%  the sharp crack for 10 layers of blunting.  The observed decrease is
%  closer to 4\%, probably because the corner is not infinitely sharp
%  due to the discreteness of the lattice.}
Clearly other effects are important.

%%%%%%%%
%
% FIGURE 7
%
%%%%%%%%
\begin{figure}[t]
  \begin{center}
    \epsfig{file=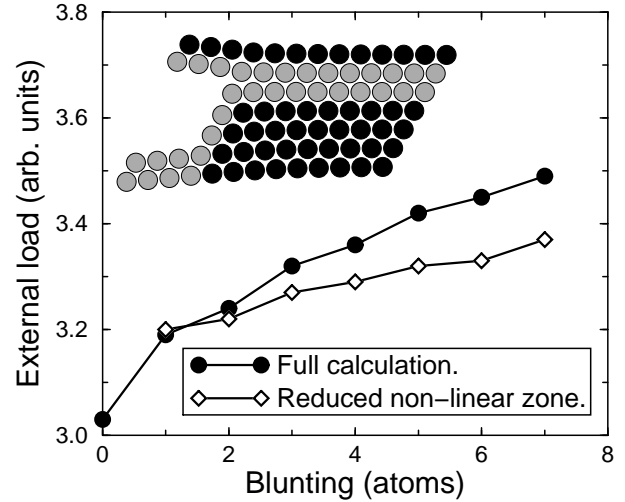,clip=}
    \vspace{2mm}
    \caption{The effect of reducing the non-linear zone at the crack
      tip.  When the atoms shown in black are removed from the
      non-linear zone (i.e.\ they are restricted to interact through
      linear forces), the effect of the blunting is reduced.  The
      range parameter in the force law is $l = 0.25$.}
    \label{fig:inelastic}
  \end{center}
\end{figure}
One such effect is the non-linearities in the force law.  We have
found that these nonlinearities explain the increase in the cleavage
stress intensity seen in figure \ref{fig:numresults}.  To evaluate
their effects, we have partly ``turned them off'' by reducing the size
of the non-linear zone.  The results are illustrated in figure
\ref{fig:inelastic} for $l = 0.25$.  Here the non-linear bonds (cf.\ 
inset) include only the bonds between surface atoms.  Reducing the
size of the non-linear zone does not change the critical loading when
the blunting is only one atomic layer.  For larger values of the
blunting, reducing the non-linear zone reduces the critical load to
cleave.  The effect of the blunting the crack beyond one layer thus
becomes smaller when the non-linear zone is reduced.  This is caused
by a stretching of the bonds of the atoms near the end of the crack.
The non-linear force-law permits bonds to be stretched further than
would otherwise be the case, thus accommodating more of the strain.
This will tend to unload the crack tip bond, inhibiting cleavage.

This leads us to conclude that the inhibition of cracking caused by
the first layer of blunting is mainly due to a change in the elastic
fields around the crack, but that further inhibition is caused by
non-linear bonds along the end of the crack unloading the crack tip.

Since the effects of the nonlinearities can be expected to depend
somewhat on the actual potential used in the simulations, it is not
entirely clear that these effects will be present in three-dimensional
simulations using more complicated potentials.  The effect of the
changes in the elastic field are going to be preserved, so we can
expect that the effect of the first layer of blunting will be
preserved.  However, it is unclear how effects like surface stress and
surface relaxations, typically seen both in experiments and in
many-body potentials, will change the behavior as the blunting is
increased.
Here it is reassuring to note that recent simulations of cracks in
NiAl\cite{Lu95} and in Al\cite{Gu95} give results very similar to the
ones observed here.  Both sets of simulations use the Embedded Atom
many-body potentials.  The crack geometries differ somewhat from this
work, but the increases in the critical loads are similar.  For one
crack orientation the shift from cleavage to dislocation emission is
observed in NiAl, similarly to the results presented here.

It is now clear that the presence of sharp corners in the blunt cracks
is highly significant for its behavior, since stress concentrations
appear at these corners that may be almost as strong as the stress
concentration at the tip of a sharp crack.  Macroscopically blunted
cracks may appear to have a smooth shape, but when the blunting is
only a few tens of lattice constants, or even smaller, the crystal
structure makes it almost inevitable for sharp corners to be present
in the crack tip configuration.  Furthermore the dislocation emission
processes creating the blunting will almost unavoidably lead to
angular crack tip shapes similar to the one studied here, and will not
create smoothly blunted cracks.  The exact angles of the corners will,
of course, depend on the detailed crystallography.  From figure
\ref{fig:exponent} it can be inferred that for $90^\circ$ and
$120^\circ$ corners, the effect of the blunting on the stress fields
is still small, although the effect is expected to be significant at
least in the $120^\circ$ case.

The change in deformation mode, from a brittle cleavage to a ductile
emission as the crack is blunted on the atomic scale, may have a
significant impact on the macroscopic behavior of the material.
Mesarovic\cite{Me96,Me96b} has shown how a crack tip moving through a
material will attract dislocations located in a relatively broad strip
ahead of the crack tip, causing them to move towards the tip and
ultimately collide with it.  Such collisions will blunt the crack, and
in some materials be sufficient to arrest it.

The two-dimensional nature of the simulations presented here invoke
the picture of infinitely long straight dislocations interacting with
a straight crack tip.  In reality neither the crack nor the
dislocations have such idealized shapes.  The dislocations interacting
with the crack are likely to be short segments or loops, and only a
short segment of the crack is thus likely to be blunted.  The result
is a crack front where short segments are arrested while the rest of
the crack moves on, until either the local stress intensity factor
becomes sufficient to reinitiate crack propagation, or until the sharp
crack moves around the obstacle.  This local ``crack trapping effect''
gives rise to a gradually increasing crack tip toughness as the
dislocation density is increased\cite{Me96b}, instead of the complete
inhibition of sharp crack propagation that two-dimensional simulations
would suggest.

Recently, Beltz {\em et al.\/}\cite{BeRiShXi96} have proposed a
continuum plasticity model for cleavage.  They show that the
``shielding ratio'', i.e.\ the ratio between the macroscopic toughness
and the microscopic crack tip toughness, depend sensitively on the
ratio between the microscopic crack tip toughness and the yield
stress.  This means that a modest increase in the microscopic
toughness of the crack tip may give an enormous increase in the
macroscopic toughness of the material.  Mesarovic\cite{Me96b} uses
this model to show that if collisions with pre-existing dislocations
locally arrest the crack, they can increase the crack tip toughness by
a factor 5--7 causing an increase of the macroscopic toughness by 2--3
orders of magnitude.  In this way he can explain the vast difference
in fracture toughness of iron-silicon and tungsten at 77K.
%%%%Mesarovic\cite{Me96b} applies
%%%%this model to show that the vast difference in fracture toughness of
%%%%iron-silicon and tungsten at 77K could be explained by collisions
%%%%between the crack tip and pre-existing dislocations.  The difference
%%%%would be caused by the much higher dislocation density in iron-silicon
%%%%($10^{13} m^{-2}$, Ref.~\onlinecite{SaNoIm73}) compared to tungsten
%%%%($10^9 m^{-2}$, Ref.~\onlinecite{LiSh84}), combined with a higher
%%%%dislocation mobility in iron-silicon.  This gives an increase in the
%%%%rate of collisions sufficient to increase the crack tip toughness to
%%%%5--7 times the Griffith value\cite{Me96b}, in agreement with
%%%%experimental estimates\cite{CuKn78}.  According to the continuum
%%%%plasticity model, this increase in the microscopic crack tip toughness
%%%%is sufficient to give rise to the very high macroscopic toughness
%%%%observed in iron-silicon.

This mechanism, where collisions with preexisting dislocations blunt
the crack and thereby arrest segments of the crack, will only lead to
a high macroscopic toughness if the blunting arrests the crack
completely, or significantly increases the load required to propagate
the crack further.  We find that the complete arrest is a possibility,
since cracks in some materials (those with high values of $l$ in this
study) will start emitting dislocations in stead of propagating as
soon as the crack is blunted.  Other materials, however, will not
change behavior.  In these materials the effect of blunting is
negligible, and the mechanism discussed above will {\em not\/} lead to
a high toughness.

\section{The blunt crack under mode II loading.}
\label{sec:modeII}

\begin{figure}[t]
  \begin{center}
    \epsfig{file=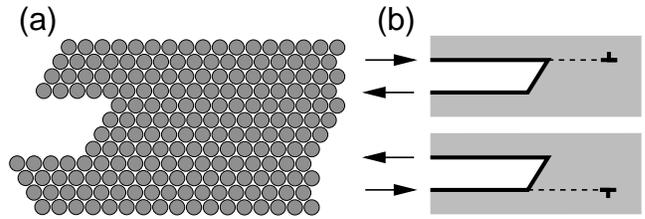,clip=}
    \vspace{2mm}
    \caption{{\em Left:\/} Geometry of the non-linear zone under mode
      II loading.  {\em Right:\/} Due to the asymmetry of the crack,
      it is necessary to keep track of the sign of the applied mode II
      load.  The loading indicated in the upper figure will be called
      {\em positive}, the loading in the lower figure {\em negative}.
      The figures also indicate how dislocations are emitted in the
      two cases, see text.}
    \label{fig:mode2geo}
  \end{center}
\end{figure}

We have also investigated the behavior of blunt cracks under mode II
loading.  Figure \ref{fig:mode2geo} shows the non-linear zone used
for these simulations.  In this case the spurs allowing dislocation
emission on inclined planes have been removed.
%%%%%%%%
%
% FIGURE 9
%
%%%%%%%%
%%%%%%%%
%
% FIGURE 10
%
%%%%%%%%
\begin{figure}[t]
  \begin{center}
    \epsfig{file=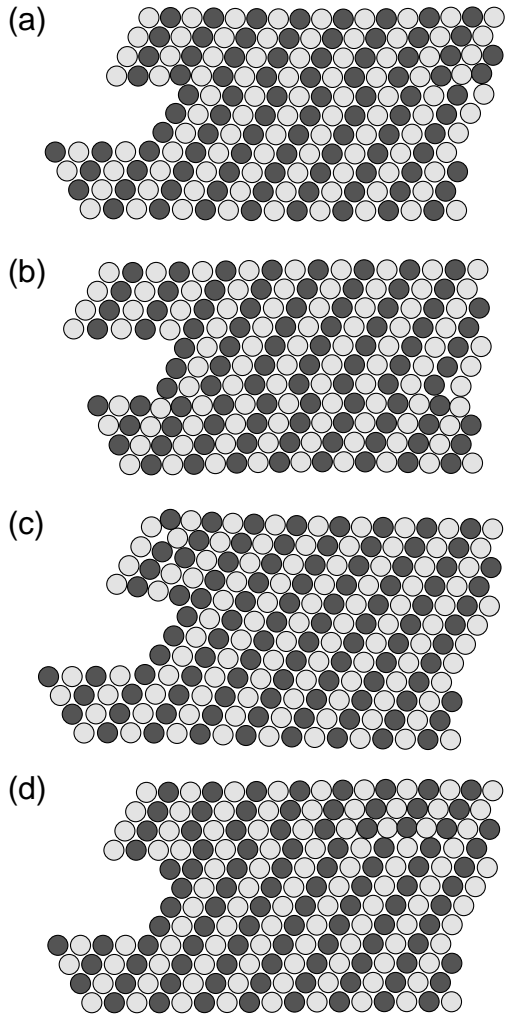,clip=}
    \vspace{2mm}
    \caption{(a) Dislocations emitted from the upper corner under positive
      load.  (b) Dislocations emitted from the lower corner under
      negative load.  (c) Collapse of the corner of the crack under
      positive mode II loading for sufficiently high blunting.  (d)
      Emission of a dislocation at a slip plane above the expected
      plane, accompanied by a reconstruction of the corner.  {\em
        Note:\/} In the two upper figures the initial blunting was
      three layers, in the lower figures it was four layers.  In all
      the figures the atoms have been given two different colors,
      forming straight lines before the dislocation was emitted.  The
      path followed by the dislocation can thus easily be seen.}
    \label{fig:mode2emit}
  \end{center}
\end{figure}
Since the crack is not symmetric, the sign of the load is important.
The sign convention used is given in figure \ref{fig:mode2geo}.  When
the crack is loaded, it is seen that dislocations are not necessarily
emitted in the sharpest corner.  Under negative load the dislocation
is emitted in the lower, more blunt corner, where the stress
concentration is smaller (see figure \ref{fig:mode2emit}).  This is an
effect of the surface energy, in both cases the dislocation is emitted
in such a way that the surface area is reduced by the emission.  The
surface energy thus assists the dislocation nucleation, and this
effect dominates that of the stress singularity.  This effect is
similar to the ``ledge effect'' under mode I loading, where the
creation of extra surface at the crack tip may dominate the energetics
of the emission\cite{ZhCaTh94}.

%%%%%%%%
%
% FIGURE 11
%
%%%%%%%%
\begin{figure}[t]
  \begin{center}
    \epsfig{file=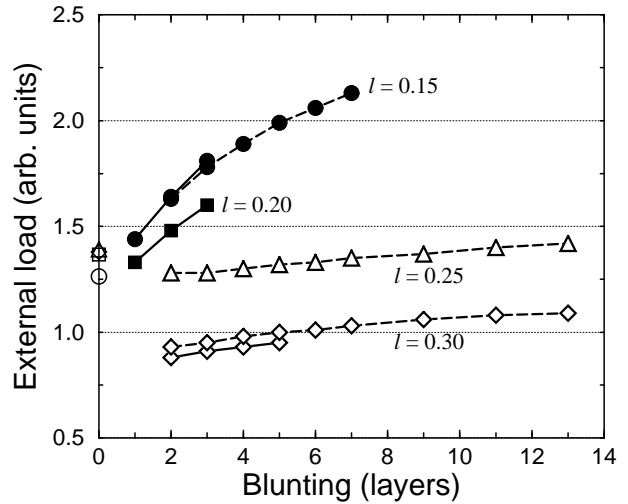,clip=}
    \vspace{2mm}
    \caption{The external load required to emit a dislocation under
      positive mode II loading, for a variety of force laws.  Symbols
      connected with a solid line are data points obtained with the
      full nonlinear zone.  Symbols connected by a dashed line are
      data points calculated with a reduced nonlinear zone.  Open
      symbols indicate a slightly different emission mode, see text.
      For most force laws, the crack collapsed if the blunting was
      only one layer, and no data could be obtained.  The symbols on
      the y-axis are the Rice prediction for the sharp crack.}
    \label{fig:mode2posresults}
  \end{center}
\end{figure}
Figure \ref{fig:mode2posresults} shows the load required for mode II
emission, under positive loading.  No measurements were attempted for
the sharp crack, since it would require a simultaneous mode I loading
to keep the crack open.  In the graph, filled symbols connected by a
solid line indicated data obtained using the full non-linear zone
shown in figure \ref{fig:mode2geo}.  For four or more layers of
blunting the corner collapsed when a load was applied, see figure
\ref{fig:mode2emit}c\cite{Endnote2}.  The collapse could be prevented
by using a smaller nonlinear zone, where the upper and lower two
layers in figure \ref{fig:mode2geo} are removed from the nonlinear
zone.  The data obtained in this way are shown connected by a dashed
line.  The reduction of the non-linear zone is seen to cause a tiny
change in the force required to emit a dislocation.  For the force law
$l = 0.25$ the corner collapsed for all levels of blunting, only data
obtained with the reduced nonlinear zone is given.  For $l = 0.30$,
dislocation emission again occurred before collapse of the corner for
sufficiently low levels of blunting.  For all force laws with $l \ge
0.25$ the emitted dislocation immediately climbed one slip plane up,
leaving an atom in the corner (figure \ref{fig:mode2emit}d).  These
cases are shown with open symbols on the graph.  They clearly have a
different dependence on the blunting, and since the dislocation moves
along the very edge of the reduced non-linear zone, such reduction has
a significant effect on the force required to emit.

%%%%%%%%
%
% FIGURE 12
%
%%%%%%%%
\begin{figure}[t]
  \begin{center}
    \epsfig{file=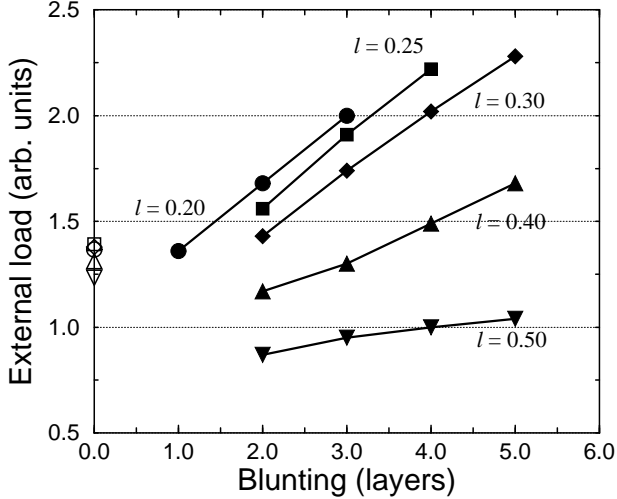,clip=}
    \vspace{2mm}
    \caption{The external load required to emit a dislocation under
      negative mode II loading, for a variety of force laws.  For most
      force laws, the crack collapsed if the blunting was only one
      layer, and no data could be obtained.  No data was obtained for
      the $l = 0.15$ force law, since the crack branched instead of
      emitting a dislocation.  The symbols on the y-axis are the Rice
      prediction for the sharp crack.}
    \label{fig:mode2negresults}
  \end{center}
\end{figure}
The results of loading the crack in the negative direction is shown in
figure \ref{fig:mode2negresults}.  Since the material is very brittle
for $l = 0.15$, the crack preferred to branch and no emission data
were obtained for that force law.

For both directions of the load, it is seen that the effect of the
blunting is far more pronounced than in the mode I case (figure
\ref{fig:numresults}).  This is in agreement with the expectations
from continuum linear elasticity, as shown in figure
\ref{fig:exponent}.  The stress singularity near the 60$^\circ$ corner
is reduced from $r^{-0.5}$ to $r^{-0.27}$, giving a significant
reduction of the stress at the corner as the blunting is increased.
For the force laws $l = 0.15$ and $l = 0.20$, the magnitude of
the effect of the blunting is consistent with this change in the
exponent, if we assume that the dislocation is emitted when the shear
stress one lattice constant ahead of the crack tip reaches a critical
value.  For the other force laws the unusual emission process
mentioned above gives a different dependence of the blunting.

For negative loading, emission is seen from the 120$^\circ$ corner.  The
dependence on blunting is even stronger than for positive loading.  In
this case continuous linear elasticity predicts {\em no\/} stress
singularity in the corner, and indeed no stress at the corner,
provided the load is pure mode II.  However, the symmetry axes of the
corners are rotated with respect to the load, so in neither case is
the load pure mode II as assumed in the Appendix, but
is a mixture of mode I and mode II.

For both signs of loading, the effect of the surface energy is clearly
seen by comparing with the Rice predictions\cite{Ri92} for the
sharp cracks, as modified to reflect the plane stress situation
studied\cite{ZhCaTh93,Zh93} (the original criterion was developed for
plane strain)
\begin{equation}
  \label{rice}
  K_{IIe} = \sqrt{2 \gamma_{us} \mu (1 + \nu)}
\end{equation}
where $\gamma_{us}$ is the so-called ``unstable stacking fault
energy''\cite{Ri92}.  In figures \ref{fig:mode2posresults} and
\ref{fig:mode2negresults} the Rice predictions are plotted along the
y-axes.  These predictions have been shown to be in good agreement
with atomic simulations of sharp cracks\cite{ZhCaTh93}, and predict
only a small variation of the critical load for emission when the
force law is varied, since $\gamma_{us}$ does not vary much for the
values of $l$ studied here.  The far larger observed variation of the
critical load is caused by the surface energy, and it is seen that a
high surface energy (high value of $l$) enhances the emission
significantly.  The effect of the surface energy on the emission
process will depend on the ratio between the surface energy and the
unstable stacking fault energy ($\gamma_s/\gamma_{us}$), since
$\gamma_{us}$ controls the energy required to create the dislocation.
For most fcc metals it is approximately 5--10, for bcc metals
1--7\cite{Ri92}; in our simulation it varies from 1.4 ($l = 0.15$) to
4.3 ($l = 0.50$).  We would thus expect that the effect of the surface
energy will be at least as significant in real materials as in our
experiments.

As is seen in figure \ref{fig:exponent},  the stress singularity in
mode III is stronger that in mode II, but weaker than in mode I.  We
would therefore expect that a blunt crack loaded in mode III behaves
in a way that is intermediate between the observed behaviors of mode I
and mode II cracks, i.e. we would expect a dependence on the critical
load on the blunting that is stronger than for mode I, but weaker than
for mode II.  Since the surface area is not changed under mode III
emission, we would not expect the strong dependence on the surface
energy that we observe in mode II.

\section{Conclusion.}
\label{sec:conclusion}

We have examined the behavior of blunted cracks under mode I and mode
II loading.  The mode I results indicate a new mechanism for enhanced
ductility in some materials.  We find that blunting the crack
increases the force required to propagate the crack, but only by a
surprisingly modest amount.  Ten layers of blunting causes an increase
of 15--20\%.  The relative modest magnitude of the effect can be
attributed to the sharp corner of the blunt crack, where the stress
singularity under mode I loading is almost as strong as for a sharp
crack.  Despite the small effect on the cleavage criterion, for many
force laws the blunting causes the crack to change behavior and to
start emitting dislocations instead of propagating.  This has the
consequence that if a crack in such a material absorbs a dislocation
at the crack tip, a segment of the crack will locally be arrested,
leading to an increase in the microscopic crack tip toughness, which
further may lead to a much larger increase in the macroscopic
toughness of the material.  The blunting effects presented here may
thus, in combination with the effect of attracting dislocations to the
crack tip described by Mesarovic\cite{Me96,Me96b}, cause an increase
in the fracture toughness of materials by many orders of magnitude.

Under mode II loading dislocation emission always occurs in such a
way as to decrease the total surface area of the crack.  The surface
energy will thus aid the emission process, and may dominate over the
stress singularity.  The emission is thus not necessarily from the
corner with the larger stress concentration.  When the surface energy
is high, the emission is seen to occur at loads significantly below
the value predicted by Rice.  The effects of blunting are greater than
for mode I loading, in agreement with the predictions of linear
elasticity.

%\section{Acknowledgements}
%\label{sec:ack}
\acknowledgements

We would like to thank Robb Thomson, Rob Phillips and Vijay Shastry
for many fruitful discussions, and for reading the manuscript.  This
work was supported by the Department of Energy under Grant Number
DE-FG02-84ER45130, by the National Institute of Standards and
Technology under award 60NANB4D1587, and by the Office of Naval
Research under Grant Number N00014-92-J-4049.

\appendix

\section*{Analytic treatment of the blunt crack.}
%\label{sec:analytic}

In order to better understand the simulations, we have investigated the
behavior of the stress field near a blunt crack, in the framework of
linear elasticity.  Unfortunately, an exact analytic solution for the
stress field around a blunt crack loaded in mode I or II can only be
found for an elliptic crack in an infinite medium\cite{In13}.  In this
section, we show that for some geometries the problem can be solved
under the assumption of antiplane strain (i.e.\ mode III loading).  We
then use simpler geometries to elucidate how the change to plane
elasticity (mode I and II) will modify the results.

\subsection{The blunt crack in antiplane strain.}
\label{sec:antiplane}

\begin{figure}[t]
  \begin{center}
    \epsfig{file=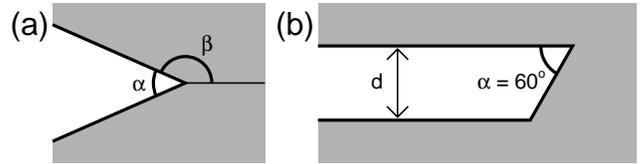,clip=}
    \vspace{2mm}
    \caption{{\em Left\/}: A wedge crack with opening angle $\alpha$.  The
      grey area is the material.  {\em Right\/}:  The geometry of a
      blunt crack.}
    \label{fig:geometry}
  \end{center}
\end{figure}

The stress field around a wedge-shaped crack loaded in mode III (see
fig \ref{fig:geometry}a) is well known\cite{OhChTh85,Th86}.  The
stress field displays a singularity at the crack tip, but the
singularity is weaker than for the sharp crack, and vanishes as the
opening angle $\alpha$ reaches $\pi$.
\begin{equation}
  \label{eq:modeIIIwedge}
  \sigma \sim r^{-p} ,\qquad p = {\pi - \alpha \over 2 \pi - \alpha}.
\end{equation}
%%%%%%%%
%
% FIGURE 13
%
%%%%%%%%

The real geometry of a blunt crack is more like the one shown in
figure \ref{fig:geometry}b.  Here we would expect the field far from
the tip to be unaffected by the blunting, i.e.\ to scale like $\sigma
\sim r^{-1/2}$.  Near the tip the field must be similar to the field
from a wedge crack, i.e. $\sigma \sim r^p$, where $p$ is given in
equation \ref{eq:modeIIIwedge}.  A weaker singularity will also be
present at the lower corner of the crack.  We can then approximate the
total stress field by a field made up of these parts, by matching the
stresses from the different solutions at an ``appropriate'' distance
from the crack tip.  It can be expected that this matching distance
will be of the order of $d$.  We will verify this na{\"\i}ve picture with
an exact solution of the problem, using conformal mappings.

In antiplane strain, the stresses can be written as the derivatives of
an analytic function, $\sigma(z) = \sigma_{yz} + i \sigma_{xz} = 2
\eta'(z)$ where $z = x + iy$.  This representation has the nice
property, that the boundary conditions (traction free surfaces) are
preserved under a conformal mapping.  If we know $\tilde{\eta}(\xi)$
is a solution to a known problem ($\Omega_\xi$), and if the function
$z = z(\xi)$ maps the known problem into a new problem ($\Omega_z$) in
such a way that the boundary maps onto the boundary, and the material
region maps onto the material region, then $\eta(z) =
\tilde{\eta}(\xi(z))$ is a solution to the new problem (provided that
$z(\xi)$ is analytic everywhere inside the material)\cite{Th86}.

We will choose the upper half plane as our known problem, and use the
Schwarz-Christoffel (S-C) transformation to generate the map.  Taking the angles as
$\pi/3$ and $2 \pi/3$ a possible S-C transformation is
\begin{equation}
  \label{eq:sctransform}
  z(\xi) = A \int d\xi\, \xi^{2/3} (\xi-1)^{1/3} + B
\end{equation}
where $A$ and $B$ are complex constants to be determined later to give
the right size, orientation and position of the blunt crack.  Using
the transformation $t = \bigl( (\xi - 1)/\xi \bigr)^{1/3}$, and
decomposing the resultant rational function into partial fractions, we
get

\begin{mathletters}
  \label{eq:transform}
  \begin{equation}
    z(\xi) = - {1 \over \pi} \sum_{k \in \{1, a, \bar{a}\}} \left[ k \,
      \mbox{Log} (t - k) + {1 \over 2 \left( t - k \right)^2} \right],
  \end{equation}
  \begin{equation}
    \qquad t = \left( {\xi - 1 \over \xi} \right)^{1/3}
  \end{equation}
\end{mathletters}
where we have set $A$ and $B$ to $9/\pi$ and 0, respectively, to get a
blunt crack of width 1 with the crack tip at the origin.  The
constants $a$ and $\bar{a}$ are equal to $\exp\left( \pm {2 \over 3}
  \pi i \right)$, i.e.\ the $k$-sum is over the roots of $k^3 = 1$.

In $\Omega_\xi$ (the upper half plane) the following solutions are
consistent with the boundary condition, traction free surfaces:
\begin{equation}
  \label{uppersolutions}
  \tilde{\eta}_n(\xi) = c_n \, \xi^n, \qquad n \in {\cal{Z}} \, \backslash
  \, \{ 0 \}, \qquad c_n \in {\cal{C}}.
\end{equation}
Each gives a solution to the blunt crack, $\eta_n(z) =
\tilde{\eta}_n(\xi(z))$.  However, we must discard all solutions with
$n \geq 2$ because the stress would diverge at infinity, and all
solutions with $n \leq -1$ because the displacement diverges at the
origin.  This leaves us $n = 1$ ($n = 0$ is not an option, since it
gives a constant $\eta$ and thus no stresses anywhere).

We will now expand the solution in the near and far field
approximations.  In both cases it is important to keep the branch cuts
of the logarithms and the non-integral powers out of the material
region, and to avoid moving between different branches of these
complex functions.

For the far field, i.e.\ for $z \rightarrow \infty$ and $\xi
\rightarrow \infty$ we get
\begin{equation}
  \label{eq:farapprox}
  z(\xi) \simeq - {9 \over 2 \pi} \xi^2 \quad \Rightarrow \quad
  \xi(z) \simeq i {\sqrt{2 \pi} \over 3} \sqrt{z},
\end{equation}
and the elastic field becomes
\begin{equation}
  \label{eq:farfield}
  \eta(z) = c \xi(z) \equiv K_{III} \sqrt{{z \over 2 \pi}}, \qquad
  \sigma(z) = { K_{III} \over \sqrt{2 \pi}} {1 \over \sqrt{z}}.
\end{equation}

The far field solution is thus identical to the stress field around a
sharp crack.  From this expansion we get the constant $c = - {3 i
  \over 2 \pi} K_{III}$.

The field near the crack tip ($z \rightarrow 0$ and $\xi \rightarrow
0$) can be found by expanding equation (\ref{eq:transform}) to fifth
order in $(-z)^{1/3}$ (It vanishes to fourth order!).  We then get
\begin{mathletters}
  \label{eq:nearapprox}
  \begin{eqnarray}
    z(\xi) &\simeq& e^{i \pi} \left( {27 \over 5 \pi} \right)
    (-\xi)^{5/3}, \\
    \xi(z) &\simeq& - e^{-3 \pi i / 5} \left( {5 \pi \over 27}
    \right)^{3/5} z^{3/5}
  \end{eqnarray}
\end{mathletters}
and get the stress field
\begin{equation}
  \label{eq:nearfield}
  \sigma(z) = e^{-\pi i / 10} K_{III} \gamma z^{-2/5}, \qquad \gamma =
  {3^{1/5} \over (5 \pi)^{2/5}}.
\end{equation}

We see that we find the same behavior as near a wedge crack with the
same opening angle.

\begin{figure}[t]
  \begin{center}
    \epsfig{file=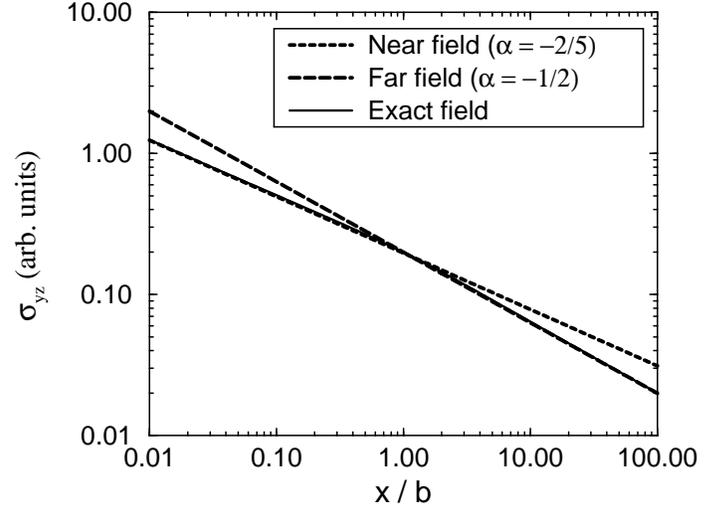,clip=}
    \vspace{2mm}
    \caption{The stress field ahead of a blunt crack, loaded in mode III.}
    \label{fig:mode3stress}
  \end{center}
\end{figure}

The exact solution is
\begin{equation}
  \label{eq:exactfield}
  \sigma(z) = - i {3 \over \pi} K_{III} \xi'(z)
\end{equation}
where $\xi'(z)$ must be found numerically, since equation
(\ref{eq:transform}) cannot be inverted analytically.  Figure
\ref{fig:mode3stress} compares the exact solution to the asymptotic
expressions; it is seen that the asymptotic approximations are quite
good everywhere on the $x$ axis.
%%%%%%%%%
%
% FIGURE 14
%
%%%%%%%%%

\subsection{The blunt crack in plane elasticity.}
\label{sec:planeblunt}

The field equations of plane elasticity (plane stress and plane
strain) do not allow the same elegant treatment as in antiplane
strain.  Although a conformal mapping technique is
possible\cite{Th86}, the boundary conditions are not preserved under
the mapping transformation, complicating the process considerably.  In
many cases such an approach can still be fruitful, especially if
combined with the integral method of Muskhelishvili\cite{Mu53,TiGo70},
but the sharp corners of the blunt crack causes unwieldy singularities
to appear.  If the conformal mapping is expanded as a (truncated)
power series, the problem of a blunt crack is soluble using
Muskhelishvili's integral method\cite{Ku96priv}, but the resultant
series expression of the stress fields can generally not be summed
analytically.

Whereas the blunt crack cannot be solved exactly, a solution for the
wedge crack can be found, using a different approach.  We describe the
stress state of the system by its Airy stress
function\cite{TiGo70,Ai1862} $\Phi$.  The stress function must satisfy
the biharmonic equation $\nabla^4 \Phi = 0$, and the stresses are then
given as suitable derivatives of $\Phi$.

We are interested in solutions to the biharmonic equation that fall
off as a power law.  In polar coordinates such solutions have the form
\begin{equation}
  \Phi(r,\theta) = r^{\lambda + 1} f(\theta)
\end{equation}
where
\begin{eqnarray}
  f_\lambda(\theta) &=& c_1 \sin (\lambda + 1 ) \theta + c_2 \cos
  (\lambda +1) \theta \nonumber \\
  && + c_3 \sin (\lambda - 1) \theta + c_4 \cos
  (\lambda - 1) \theta
\end{eqnarray}
giving the stresses
\begin{mathletters}
\begin{eqnarray}
  \sigma_{rr} &=& {1 \over r} {\partial \Phi \over \partial r} + {1
    \over r^2} {\partial^2 \Phi \over \partial \theta^2} \nonumber \\
  &=&  r^{\lambda
    - 1} \left( f_\lambda''(\theta) + (\lambda+1) f_\lambda(\theta)
  \right),\\ 
  \sigma_{\theta\theta} &=& {\partial^2 \Phi \over \partial r^2} =
  r^{\lambda-1} \lambda (\lambda+1) f_\lambda(\theta),\\
  \sigma_{r\theta} &=& - {\partial \over
    \partial r} \left( {1 \over 
      r} {\partial \Phi \over \partial \theta} \right) = -
  r^{\lambda-1} \lambda f_\lambda'(\theta).
\end{eqnarray}
\end{mathletters}
We now demand traction-free surfaces, i.e.\ $\sigma_{\theta\theta} =
\sigma_{r\theta} = 0$ for $\theta = \pm \beta$ where $\beta$ is given
by the opening angle (figure \ref{fig:geometry}): $\beta = \pi -
\alpha / 2$.  We thus demand that $f(\pm \beta) = f'(\pm \beta) = 0$,
giving
\begin{mathletters}
  \begin{equation}
    c_2 \cos \bigl( (\lambda+1) \beta\bigr) + c_4 \cos \bigl(
    (\lambda-1)\beta \bigr) = 0 \label{eq:lineqI},
  \end{equation}
  \begin{equation}
    c_2 (\lambda+1) \sin \bigl( (\lambda+1)\beta \bigr) +
    c_4 (\lambda-1) \sin \bigl( (\lambda-1)\beta \bigr) = 0
    \label{eq:lineqII},
  \end{equation}
  \begin{equation}
    c_1 \sin \bigl( (\lambda+1) \bigr) + c_3 \sin \bigr(
    (\lambda-1)\beta \bigr) = 0 \label{eq:lineqIII},
  \end{equation}
 \begin{equation}
    c_1 (\lambda+1) \cos \bigl( (\lambda+1)\beta \bigr) + 
    c_3 (\lambda-1) \cos \bigl( (\lambda-1)\beta \bigr) = 0.
    \label{eq:lineqIV}
  \end{equation}
\end{mathletters}
%\begin{mathletters}
%  \begin{eqnarray}
%    c_2 \cos \bigl( (\lambda+1) \beta\bigr) + c_4 \cos \bigl(
%    (\lambda-1)\beta \bigr) &=& 0 \label{eq:lineqI},\\
%%
%    c_2 (\lambda+1) \sin \bigl( (\lambda+1)\beta \bigr) + && \nonumber
%    \\
%%
%    c_4 (\lambda-1) \sin \bigl( (\lambda-1)\beta \bigr) &=& 0
%    \label{eq:lineqII},\\
%%
%    c_1 \sin \bigl( (\lambda+1) \bigr) + c_3 \sin \bigr(
%    (\lambda-1)\beta \bigr) &=& 0 \label{eq:lineqIII},\\
%%
%    c_1 (\lambda+1) \cos \bigl(
%    (\lambda+1)\beta \bigr) + && \nonumber \\
%%
%    c_3 (\lambda-1) \cos \bigl( (\lambda-1)\beta \bigr) &=& 0.
%    \label{eq:lineqIV}
%  \end{eqnarray}
%\end{mathletters}

These two sets of two homogeneous linear equations only have
non-trivial solutions if one of the determinants vanish.  For the
first two equations this happens for
\begin{equation}
  \label{eq:eigenI}
  \lambda \sin(2 \beta) + \sin (2 \lambda \beta) = 0
\end{equation}
and we get the stress fields
\begin{mathletters}
  \begin{eqnarray}
    \label{eq:modeIstress}
    \sigma_{rr} &=& c_2 r^{\lambda-1} \lambda \Bigl( -(\lambda+1) \cos
    \bigl[ (\lambda+1)\theta \bigr] \nonumber \\
    && \phantom{c_2 r^{\lambda-1} \lambda \Bigl(} + (\lambda-3) Q_A
    \cos \bigl[ (\lambda-1)\theta \bigr] \Bigr),\\ 
    \sigma_{\theta\theta} &=& c_2 r^{\lambda-1} \lambda(\lambda+1)
    \Bigl( \cos \bigl[ (\lambda+1)\theta \bigr] \nonumber \\
    && \phantom{c_2 r^{\lambda-1} \lambda(\lambda+1) \Bigl(}
    - Q_A \cos \bigl[ (\lambda-1)\theta \bigr] \Bigr),\\ 
    \sigma_{r\theta} &=& c_2 r^{\lambda-1} \lambda \Bigl( (\lambda+1) 
    \sin \bigl[ (\lambda+1)\theta \bigr] \nonumber \\
    && \phantom{ c_2 r^{\lambda-1} \lambda \Bigl( }- (\lambda-1) Q_A \sin
    \bigl[ (\lambda-1)\theta \bigr] \Bigr),
  \end{eqnarray}
\end{mathletters}
where
\begin{equation}
  \label{eq:QA}
  Q_A = -{c_4 \over c_2} = {\cos \bigl[(\lambda+1)\beta \bigr] \over
    \cos \bigl[ (\lambda-1)\beta \bigr]}.
\end{equation}
The solutions of equation (\ref{eq:eigenI}) are constrained by the requirement
that the stresses vanish at infinity and that the displacements remain
finite at the crack tip: $0 \le \lambda < 1$.  For the sharp crack
($\alpha = 0$ and $\beta = \pi$) we get $\lambda = 1/2$ and find the
known stress field of the crack under mode I loading\cite{Th86}.

Equations (\ref{eq:lineqIII}) and (\ref{eq:lineqIV}) give another
eigenequation:
\begin{equation}
  \label{eq:eigenII}
  \lambda \sin(2\beta) - \sin(2\lambda\beta) = 0
\end{equation}
giving the stresses
\begin{mathletters}
  \label{modeIIstress}
  \begin{eqnarray}
    \sigma_{rr} &=& c_1 r^{\lambda-1} \lambda \Bigl( -(\lambda+1) \sin
    \bigl[ (\lambda+1)\theta \bigr] \nonumber \\
    && \phantom{c_1 r^{\lambda-1} \lambda \Bigl(}
    + (\lambda-3) Q_B \sin \bigl[ (\lambda-1)\theta \bigr] \Bigr), \\
    \sigma_{\theta\theta} &=& c_1 r^{\lambda-1} \lambda(\lambda+1)
    \Bigl( \sin \bigl[ (\lambda+1)\theta \bigr] \nonumber \\
    && \phantom{c_1 r^{\lambda-1} \lambda(\lambda+1) \Bigl(}
    - Q_B \sin \bigl[ (\lambda-1)\theta \bigr] \Bigr), \\
    \sigma_{r\theta} &=& - c_1
    r^{\lambda-1} \lambda \Bigl( (\lambda+1) 
    \cos \bigl[ (\lambda+1)\theta \bigr] \nonumber \\
    && \phantom{- c_1 r^{\lambda-1} \lambda \Bigl(}
    + (\lambda-1) Q_B \cos \bigl[(\lambda-1)\theta \bigr] \Bigr),
  \end{eqnarray}
\end{mathletters}
where
\begin{equation}
  \label{eq:QB}
  Q_B = - {c_3 \over c_1} = {\sin \bigl[(\lambda+1)\beta\bigr] \over
    \sin \bigl[ (\lambda-1)\beta \bigr] }.
\end{equation}
Again $\lambda$ is constrained to $0 \le \lambda < 1$, which can only
be fulfilled for opening angles less than $\alpha < 103^\circ$.
For the sharp crack we recover the mode II solution.

Figure \ref{fig:exponent} shows the exponent ($\lambda - 1$) of the
stress singularity as a function of the opening angle for wedge cracks
loaded in mode I, II and III.

The behavior of the mode II solution at $\alpha = 103^\circ$ is
peculiar.  As the angle approaches $180^\circ$ the mode I solution
approaches a half-plane under uniform tensile stress.  Such a behavior
is not possible in mode II, since the boundary conditions prevent a
uniformly sheared half-plane (an attempt to apply shear alone would
only result in a rotation).  It is thus clear that the stress must go
to zero as the opening angle approaches $180^\circ$.  However that
behavior is seen already at $103^\circ$.  Between $103^\circ$ and
$180^\circ$ a stress field that decays as a power law in $r$ is not
possible, nor is a stress constant in $r$.  It is hard to imagine
another form of the stress where it increases as $r$ decreases, since
any other form would require that another length scale enters, and
linear elasticity is known to be scale-invariant.

As a curiosity it can be mentioned that this peculiar behavior also
has been seen in a related problem: In the beginning of the century
the problem of a wedge loaded at its apex by a couple of moment $M$
was solved\cite{Ca12,In22}.  A solution was found with the resulting
stress field decaying as $r^{-2}$, diverging for (in our notation)
$\alpha = 103^\circ$.  This strange singular behavior at a physically
meaningful angle was later investigated by Sternberg and
Koiter\cite{StKo58}.  They replaced the couple by a physically
realizable force distribution of size $a$, and then solved the stress
field for $r \gg a$.  They found that for $\alpha > 103^\circ$ the
solution converged to the original solution, but below $103^\circ$ an
extra (dominating) term appeared, making the concept of a couple
loading unphysical.  That extra term is identical to the mode II
solution above!

%\bibliographystyle{prsty}
%\bibliography{local}

\begin{thebibliography}{10}
\bibitem{Endnote1}
The change in the exponent from $-1/2$ to $-0.488$ would indicate that the
  strain five atoms ahead of the crack should have decreased by approximately
  1\% compared to the sharp crack for 10 layers of blunting. The observed
  decrease is closer to 4\%. This difference is due to mode mixing, the
  $30^\circ$ misorientation of the crack introduces some mode II loading. The
  exponent for the mode II part is $-0.269$, The blunting thus has a larger
  effect on this part of the strain.
\bibitem{Endnote2}
This collapse is actually emission of a dislocation on another slip plane. Such
  emission processes are not studied further in this paper.
\end{thebibliography}

%
%  Figures come here
%

%%%%%%
%
%  FIGURE 1
%
%%%%%%
%%%%%%%%%
%
% FIGURE 2
%
%%%%%%%%%

%%%%%%%%%
%
% FIGURE 3
%
%%%%%%%%%

%%%%%%%
%
% FIGURE 4
%
%%%%%%%

%%%%%%%
%
% FIGURE 5
%
%%%%%%%

%%%%%%%
%
% FIGURE 6
%
%%%%%%%

%%%%%%%%
%
% FIGURE 7
%
%%%%%%%%

%%%%%%%%%%%
%%%%%
%%%%% FIGURE 8
%%%%%
%%%%%%%%%%%
%%%%\begin{figure}[t]
%%%%  \begin{center}
%%%%%    \epsfig{file=interface.ps}
%%%%%    \vspace{2mm}
%%%%    \caption{A crack at an interface.  Emission only occurs in
%%%%      the downwards direction, possibly because this direction is
%%%%      crystallographically favored, or because the two materials have
%%%%      different properties.  In (a) all dislocations are emitted on
%%%%      the same slip plane, and the crack tip is seen leaving the
%%%%      interface.  In (b) the dislocations are emitted on adjacent
%%%%      planes, but the crack tip stays on the interface.  The cracks
%%%%      simulated in this paper emit in a way that would result in the
%%%%      second behavior.}
%%%%    \label{fig:interface}
%%%%  \end{center}
%%%%\end{figure}

%%%%%%%%
%
% FIGURE 9
%
%%%%%%%%

%%%%%%%%
%
% FIGURE 10
%
%%%%%%%%

%%%%%%%%
%
% FIGURE 11
%
%%%%%%%%

%%%%%%%%
%
% FIGURE 12
%
%%%%%%%%

%%%%%%%%
%
% FIGURE 13
%
%%%%%%%%

%%%%%%%%%
%
% FIGURE 14
%
%%%%%%%%%

\end{document}